\documentclass[11pt]{article}
\usepackage{epsf}
\usepackage{amsmath}
\usepackage{graphics}
\usepackage{cite}

\renewcommand{\thefootnote}{\#\arabic{footnote}}
\begin{document}

\newcommand{\gtrsim}{ \mathop{}_{\textstyle \sim}^{\textstyle >} }
\newcommand{\lesssim}{ \mathop{}_{\textstyle \sim}^{\textstyle <} }

\newcommand{\rem}[1]{{\bf #1}}

\renewcommand{\thefootnote}{\fnsymbol{footnote}}
\setcounter{footnote}{0}
\begin{titlepage}

\def\thefootnote{\fnsymbol{footnote}}

\hfill April 2018\\
\vskip .5in
\bigskip
\bigskip

\begin{center}
{\Large \bf  On the Origin and Nature of Dark Matter}

\vskip .45in

{\bf P.H. Frampton\footnote{email: paul.h.frampton@gmail.com}}

\bigskip

{Brasenose College, Oxford OX1 4AJ, UK}

\end{center}

\bigskip
\bigskip
\bigskip

\begin{abstract}

\noindent
It is discussed how the ideas of entropy and the second law of thermodynamics, 
conceived long ago during the nineteenth century, underly 
why cosmological dark matter exists and originated in the first three years of
the universe in the form of primordial black holes, a very large number of
which have many solar
masses including up to the supermassive black holes at the centres
of galaxies. Certain upper 
bounds on dark astrophysical objects with many solar masses 
based on analysis of the CMB spectrum and published in the 
literature are criticised. For
completeness we 
discuss WIMPs and axions which are leading particle 
theory candidates for the constituents of dark matter.
The PIMBHs (Primordial Intermediate Mass Black Holes) with many solar masses
should be readily detectable in microlensing experiments 
which search the Magallenic Clouds and measure
light curves with durations of from one year up to several years.
\end{abstract}

\end{titlepage}

\renewcommand{\thepage}{\arabic{page}}
\setcounter{page}{1}
\renewcommand{\thefootnote}{\#\arabic{footnote}}

\newpage

\section*{Introduction}

\bigskip

\noindent
We assume Newton's universal law for the gravitational force $F$ between every
two point particles
\begin{equation}
F = \left( \frac{G M_1 M_2}{R^2} \right)
\label{Newton}
\end{equation}

\noindent
where $G$ is a constant, $M_1, M_2$ are the two masses and $R$ is the separation,
to be valid at the scale of galaxies and clusters of galaxies. 
Then there is overwhelming observational evidence
\cite{Zwicky,Rubin} for the existence of dark matter which does not radiate electromagnetically
but which, by assumption, interacts gravitationally according to Eq.(\ref{Newton}).
The mass of such dark matter in the Milky Way is five or six times the mass of the
luminous matter which does radiate electromagnetically.

\bigskip

\noindent
There is no doubt that the dark matter exists astrophysically and cosmologically
but there is no reliable evidence for dark matter in terrestrial experiments. In the
present article we shall first address (Section 1) the reason why dark matter
exists by using thermodynamic and entropic arguments especially
the second law of thermodynamics. The use of entropy may be
unfamiliar to particle theorists because one cannot define the entropy for a 
single particle. Nevertheless, the reason why dark matter exists pre-dates
quantum field theory.
We next address (Section 2) the
nature of the dark matter and will first briefly discuss within particle theory 
WIMPs and Axions as candidates for dark matter constituents.
These are alternatives to the notion that galactic dark
matter is comprised of Primordial Intermediate-Mass Black Holes (PIMBHs).
Primordial means that they were formed in the first three years of the
expanding universe compared to the 
first three minutes for luminous matter.

\bigskip

\noindent
For the present discussion, the PIMBHs of special interest, being the most
susceptible to observational discovery by microlensing in the near future, have masses in the
narrow range
\begin{equation}
25 M_{\odot} \leq  M_{PIMBH} \leq 2,500 M_{\odot},
\label{PIMBHmass}
\end{equation}
\noindent
which is a range accessible to the microlensing experiments
with light curves less than ten years in duration for
lensing of stars in the Magellanic Clouds. Masses above the upper end
in Eq.(\ref{PIMBHmass}) up to $10^6 M_{\odot}$ are permitted
in the Milky Way halo where there is a mass cut-off caused by
disk stability\cite{XuOstriker} for MACHOS
away from the galactic centre. Near to the centre
of gravity  of the Milky Way is the
supermassive black hole SagA* with 
\begin{equation}
M_{SagA*} \simeq 4 \times 10^6 M_{\odot}
\label{MassSagA*}
\end{equation}
\noindent
and elsewhere in the visible universe there are
primordial black holes with masses anywhere up to $10^{12}M_{\odot}$
which include all the supermassive black holes at galactic centres.

\bigskip

\noindent
Black holes mean the electrically neutral objects predicted by general relativity
of Kerr type \cite{Kerr}, specified completely by mass and spin.
Having discussed both the why (Section 1) and what (Section 2)
of dark matter, we end with a Discussion section.

\newpage

\bigskip

\section{On the Origin of Dark Matter}

\bigskip

\noindent
The reason that dark matter exists is based on the thermodynamic
concepts of entropy and the second law of thermodynamics. The present
section will contain introductory discussions about entropy which was
a major accomplishment of nineteenth
century theoretical physics. Physicists studying particle theory normally use
the language of quantum field theory and can be skeptical 
about the usefulness of entropy. 
\bigskip

\noindent
Dark matter should be regarded as an astrophysical
phenomenon and its appearance in galaxies and clusters of
galaxies the result of dynamical evolution in the early universe
of extremely large
numbers of particles. This leads naturally to the employment
of the methods of statistical mechanics and thermodynamics.
To understand the cosmological dark matter,
we must therefore study assiduously Boltzmann rather than Maxwell.

\bigskip

\noindent
For the expansion of the visible universe, 
we are dealing with $\sim 10^{80}$ particles, far
more than Avagadro's number $6.023 \times  10^{23}$ molecules
per mole and therefore
a statistical treatment should give reliable predictions. Application
of entropy to the visible universe is secure because, assuming
a finite universe, it is 
a thermodynamical isolated system since no heat
enters or leaves and the visible universe can
be regarded as if surrounded by a perfect thermal insulator. 

\bigskip

\noindent
The universe is not as straightforward as, but not so much more
complicated than, a box with a mole of ideal gas and
elastic collisions between molecules and between molecules and
the walls, where with $6.023 \times 10^{23}$ molecules we can derive at low
density extremely accurate thermodynamic laws such as the ideal gas law.
For the universe, it is a time-dependent system and the exactly
accurate dynamics would require solving Boltzmann's transport equations
for all the particles, so we must instead use thermodynamic
arguments.

\bigskip

\noindent
The kinetic theory of gases is a misleadingly simple case 
where precise
macroscopic properties of a box of ideal gas are related
to the microscopic properties of the molecules. 
There are thermodynamic variables $P, V, T$ whose physical
significance in terms of experimental measurement is clear.
$S$, the entropy, is a state function. 

\bigskip

\noindent
The second half of the 19th century was when entropy was introduced
into physics. In the 1850s and culminating in 1865,
Clausius thought carefully about Carnot's earlier
cyclic model for a steam engine and it is Clausius who deserves 
the credit for introducing
both entropy and the second law of thermodynamics. Quite a pair of 
accomplishments! 
But the connection of entropy
to microscopic physics was first made by Boltzmann in 1872. 

\bigskip

\noindent
The French physicist who was the father of thermodynamics
and whose work started the intellectual path towards entropy
was Sadi Carnot (1796-1832).
In his 1824 book \cite{Carnot} S. Carnot
began a new field of research, thermodynamics,
and his Carnot Cycle was what later led Clausius in 1865 to the
idea of entropy.
The Carnot Cycle is a simple model which mimics the operation of
a steam engine.

\bigskip

\noindent
Rudolf Clausius (1822-1888), a German physicist and mathematician,
may justly be regarded as the father of entropy.  He was initially inspired by
the Carnot Cycle which requires that

\begin{equation}
\left( \frac{Q_H}{T_H} \right) = \left( \frac{Q_L}{T_L} \right)
\label{QoverT}
\end{equation}

\noindent
where $T_H,T_L$ are the absolute temperatures of the hot and cold
heat reservoirs and $Q_H,Q_L$ the heat absorbed and emitted respectively.
In the
presence of irreversible processes in a variant of the Carnot
Cycle one would, instead of the equality in Eq.(\ref{QoverT}), have
an inequality $(Q_H/T_H) < (Q_L/T_L)$
which gives rise to the second law.
This led Clausius to a definition\cite{Clausius}
for incremental entropy as the exact differential

\begin{equation}
dS = \left( \frac{\delta Q}{T} \right)
\label{dS}
\end{equation}

\bigskip

\noindent
near to thermal equilibrium and thence to the second law of thermodynamics
$d S \geq 0$.
We emphasise that Eq.(\ref{dS}) is appropriate only near to thermal equilibrium
because, for example, a thermally insulated box of ideal gas with an unlikely initial
condition, {\it e.g.} all the molecules in one corner, will rapidly increase its entropy to
approach thermal equilibrium despite the fact that $\delta Q \equiv 0$. Clausius
denoted entropy by $S$ in honour of {\bf S}adi Carnot. The early universe is
never near to equilibrium so that Eq.(\ref{dS}) does not apply: $\delta Q=0$ but $dS>0$.

\bigskip

\noindent
Clausius enunciated two laws as follows:\\
1. The energy of the universe is a constant.\\
2. The entropy of the universe tends to a maximum.\\
This succinct statement of the second law is perfect for use
in our Discussion section.

\bigskip

\noindent
For a closed, isolated homogeneous system in which all
processes in a cycle are reversible the closed loop
integral vanishes:

\begin{equation}
\oint dS = \oint \frac{\delta Q}{T} = 0
\label{loop}
\end{equation}

\noindent
which implies that the line integral is independent of the path
and hence that the increment $dS$
is uniquely defined as an exact differential at least proximate
to thermal equilibrium.
Therefore we have a sensible thermodynamic state function
$S$ whose partial differentiations with respect to the
thermodynamic variables permit an expression for relative
entropies in an ideal gas.

\bigskip

\noindent
Kinetic theory shows how the $P, V, T$ thermodynamic variables
can be related to the average motions of the molecules using
statistical mechanics. The question following Clausius's work
was how to relate the state function $S$ to microscopic variables?
Ludwig Boltzmann (1844-1906) was the physicist who solved
this problem. He had no
experimental evidence for molecules; this had to wait
thirty more years until the explanation of Brownian 
motion made in 1905 by Einstein\cite{BrownEinstein}
and Smoluchowski\cite{BrownSmoluchowski}.

\bigskip

\noindent
Boltzmann was in many ways a tragic figure. Few 
people were convinced of the reality of atoms and molecules
before the last year of his life. Further, his statistical, hence inexact,
second law of thermodynamics was strongly criticised by Maxwell (1831-1879)
who, although he believed in atoms and molecules, never accepted
Boltzmann's 1872 idea of an 
inexact law of physics. Boltzmann appreciated that his law was so
unlikely to be violated that it might as well have been exact.
Another severe criticism came from the distinguished
French mathematician Henri Poincar\'e (1854-1912)
who proved a rigorous recurrence theorem \cite{Poincare}
which states that all systems  
must return eventually to their original state. Boltzmann understood that
the time scale involved in Poincar\'e recurrence is far too long to be
physically relevant. In any case, Boltzmann's lack of recognition in the
physics and mathematics communities may have contributed in 
1906 to his suicide at the early age of 62.

\bigskip

\noindent
Boltzmann defined as {\it microstates} all the 
possible arrangements of microscopic variables
corresponding to a given fixed set of macroscopic
or thermodynamic variables. Let $p_i$ be the
probability that the system is in the $i th$ microstate.
Then introducing the constant
$k$ with the same units as $S$ he represented $S$ as

\begin{equation}
S = - k \Sigma_i  p_i \ln p_i
\label {microS}
\end{equation}

\noindent
He made the ergodic hypothesis that all the $p_i$ are equal

\begin{equation}
p_i = \frac{1}{\Omega} ~~~ {\rm for ~ all ~} ~i
\end{equation}

\noindent
whereupon
\begin{equation}
S = k \ln \Omega
\label{Boltzmann}
\end{equation}
where $\Omega$ is the total number of microstates. Eq. (\ref{Boltzmann})
is one of the most celebrated equations in all of physics.

\bigskip

\noindent
For an ideal gas, the maximisation of entropy $S$ means that
in the state of thermal equilibrium there is the maximum
uncertainty in the molecular motions. We can equivalently
say that there is the greatest disorder in thermal equilibrium,
and hence that entropy is a measure of disorder.

\bigskip

\noindent
The H theorem (1872)  \cite{Boltzmann} of Boltzmann is central to the physics
although even now in 2018 we are told it cannot be rigorously proved
mathematically because, at least far away from equilibrium, it is unknown
whether solutions of the Boltzmann transport equation
have sufficient analytic smoothness. Nevertheless,
the H theorem shows how starting from reversible microscopic
mechanics, one can arrive at non-reversible,
in a statistical sense,
macroscopic dynamics. It explicates
the second law of thermodynamics that the entropy of an
isolated system cannot decrease. Later in this paper, we shall
argue that the early visible universe can be regarded
as such an isolated system.

\bigskip

\noindent
The Boltzmann transport equation is a general requirement
for $f({\bf q}, {\bf p}, t)$ which is the distribution of molecules
with position ${\bf q}$ and momentum ${\bf p}$ at time $t$ and is written
as follows:

\begin{equation}
\frac{\partial f({\bf q}, {\bf p}, t)}{\partial t} +
\left( \frac{{\bf p}}{m} \right). \frac{\partial f({\bf q}, {\bf p}, t)}{\partial {\bf q}} +
{\bf F}.\frac{\partial f({\bf q}, {\bf p}, t)}{\partial {\bf p}} = \left( \frac{\partial f}{\partial t} \right)_{coll}
\label{transport}
\end{equation}

\noindent
where the RHS is
\begin{equation}
\left( \frac{\partial f}{\partial t} \right) = \bar{R} - R
\label{collisions}
\end{equation}

\noindent
in which $R$ dt d{\bf q} d{\bf p} is the number of collisions from time t to (t + dt)
with initial position {\bf q} to ({\bf q} + d{\bf q}) and initial momentum {\bf p}
to ({\bf p} +d{\bf p})
and $\bar{R}$ dt d{\bf q} d{\bf p} is the number of collisions from time t to (t + dt)
with final position {\bf q} to ({\bf q} + d{\bf q}) and final momentum {\bf p}
to ({\bf p} +d{\bf p}).

\bigskip

\noindent
Taking only $2 \rightarrow 2$ elastic collisions into account

\begin{equation}
R({\bf q},{\bf p_1}) = \int d {\bf p_2} d{\bf p_1^{'}} d {\bf p_2^{'}} 
{\cal P}_{{\bf p_1},{\bf p_2} \rightarrow {\bf p_1^{'}} {\bf p_2^{'}}}
f( {\bf q}, {\bf p_1}) f({\bf q},{\bf p_2})
\label{Rqp1}
\end{equation}

\bigskip

\noindent
while for the final state

\begin{equation}
\bar{R} ({\bf q},{\bf p_1}) = \int d {\bf p_2} d{\bf p_1^{'}} d {\bf p_2^{'}} 
{\cal P}_{{\bf p_1^{'}},{\bf p_2^{'}} \rightarrow {\bf p_1} {\bf p_2}}
f( {\bf q}, {\bf p_1^{'}}) f({\bf q},{\bf p_2^{'}})
\label{barRqp1}
\end{equation}

\bigskip

\noindent
In Eqs.(\ref{Rqp1}) and ({\ref{barRqp1}),
${\cal P}_{{\bf p_1},{\bf p_2} \rightarrow {\bf p_1^{'}} {\bf p_2^{'}}}$ 
is the probability density for going from initial
state ${\bf p_1}, {\bf p_2}$ to final state ${\bf p_1^{'}}, {\bf p_2^{'}}$ in time dt.

\bigskip

\noindent
Time-reversal symmetry for the microscopic scattering requires that
\begin{equation}
{\cal P}_{{\bf p_1},{\bf p_2} \rightarrow {\bf p_1^{'}}, {\bf p_2^{'}}}
= {\cal P}_{{\bf p_1^{'}},{\bf p_2^{'}} \rightarrow {\bf p_1}, {\bf p_2}}
\label{Treversal}
\end{equation}

\noindent
and therefore we may rewrite Eq.({\ref{collisions}) as

\begin{equation}
\left( \frac{\partial f}{\partial t} \right)
= \int d{\bf p_2} d{\bf p_1^{'}} d{\bf p_2^{'}} {\cal P}_{{\bf p_1},{\bf p_2} \rightarrow {\bf p_1^{'}}, {\bf p_2^{'}}} \left[ f({\bf p_1^{'}}) f({\bf p_2^{'}} )- f({\bf p_1}) f({\bf p_2}) \right]
\label{fdot}
\end{equation}

\bigskip

\noindent
To indentify entropy $S$ microscopically, presumably inspired by the
monotonic increase of the logarithm function, Boltzmann considered
what he called the H function ($S=-H$) defined by

\begin{equation}
H(t) = \int d{\bf p} f({\bf p}, t) \log f({\bf p}, t)
\label{Hfunction}
\end{equation}

\noindent
The time derivative of $H(t)$ is

\begin{eqnarray}
\left( \frac{ d H(t)}{d t} \right) &=& \int d{\bf p} \frac{\partial}{\partial t} \left[ f({\bf p}, t) \log f({\bf p}, t) \right] \nonumber \\
&=& \int d{\bf p} \left( \frac{\partial f({\bf p}, t)}{\partial t} \right) \left[ 1 + \log f({\bf p}, t) \right]
\label{Hdot}
\end{eqnarray}

\noindent
Substitution of Eq.(\ref{fdot}) into Eq.(\ref{Hdot}) gives

\begin{equation}
\left( \frac{ d H(t)}{d t} \right) = \int d{\bf p_1}d{\bf p_2} d{\bf p_1^{'}} d{\bf p_2^{'}} {\cal P}_{{\bf p_1},{\bf p_2} \rightarrow {\bf p_1^{'}}, {\bf p_2^{'}}} \left[ f({\bf p_1^{'}}) f({\bf p_2^{'}} )- f({\bf p_1}) f({\bf p_2}) \right] \left[ 1 + \log f({\bf p_1}) \right]
\label{Hdot2}
\end{equation}

\bigskip

\noindent
Because of the symmetry in Eq.(\ref{Hdot2}), we may freely replace $f({\bf p_1})$ by
$f({\bf p_2})$ in the logarithm, and add the result to Eq.(\ref{Hdot2}) to obtain

\begin{equation}
2 \left( \frac{ d H(t)}{d t} \right) = \int d{\bf p_1}d{\bf p_2} d{\bf p_1^{'}} d{\bf p_2^{'}} {\cal P}_{{\bf p_1},{\bf p_2} \rightarrow {\bf p_1^{'}}, {\bf p_2^{'}}} \left[ f({\bf p_1^{'}}) f({\bf p_2^{'}} )- f({\bf p_1}) f({\bf p_2}) \right] \left[ 1 + \log f({\bf p_1})f({\bf p_2}) \right]
\label{Hdot3}
\end{equation}

\noindent
Now we exploit the time-reversal-invariance of ${\cal P}$ in Eq(\ref{Treversal})
to arrive at the fascinating formula which is 
at the heart of Boltzmann's derivation:

\begin{eqnarray}
\left( \frac{ d H(t)}{d t} \right) &=& - \frac{1}{4} \int d{\bf p_1}d{\bf p_2} d{\bf p_1^{'}} d{\bf p_2^{'}} {\cal P}_{{\bf p_1},{\bf p_2} \rightarrow {\bf p_1^{'}}, {\bf p_2^{'}}} \left[ f({\bf p_1^{'}}) f({\bf p_2^{'}} )- f({\bf p_1}) f({\bf p_2}) \right] \times \nonumber \\
& & \left[ \log f({\bf p_1})f({\bf p_2}) -  \log f({\bf p_1^{'}})f({\bf p_2^{'}})\right]
\label{Hdot4}
\end{eqnarray}

\noindent
Eq.(\ref{Hdot4}) has a RHS which is negative semidefinite because
$(A - B)(\log A - \log B) \geq 0$ and therefore

\begin{equation}
\left( \frac{ d H(t)}{d t} \right) \leq 0
\label{Hdot5}
\end{equation}

\bigskip

\noindent
or, reverting to entropy $S(t) = -H(t)$, we have arrived at a microscopic derivation
of the second law that with a very large number of molecules $S$ cannot decrease. 

\bigskip

\noindent
The paper by Boltzmann in which he proved the H theorem has been
studied and criticised probably as much as any physics paper. One
interesting critique\cite{Neumann} is by Von Neumann (1903-1957),
available in translation \cite{Neumann2}. 

\bigskip

\noindent
The reason why the H theorem of Boltzmann is far more 
powerful than the infinitesimal definition
of $dS$ by Clausius is that it proves that $dS \geq 0$  for 
non-equilibrium systems assuming only the Boltzmann 
transport equation and the ergodic hypothesis,

\bigskip

\noindent
What is clear about $S(t)$ for a box of ideal gas is that with
thermal equilibrium Eq(\ref{Hdot5}) becomes an equality
and that $S(t)$ is a maximum. From the definition of $S$ in Eq.(\ref{Boltzmann})
this implies that the number of microstates corresponding to the thermally
equilibrated system is the highest, and that therefore the molecular motion
is the most disordered.

\bigskip

\noindent
The H theorem encapsulates this edifice of 19th-century knowledge sufficiently
to progress with some confidence from a box of ideal gas to the more 
interesting case of the early universe.

\bigskip

\begin{center}
- - - - - - - -
\end{center}

\bigskip

\noindent
To discuss cosmology we begin from Einstein's equations\cite{Einstein1,Einstein2}
of general relativity

\begin{equation}
R_{\mu\nu} - \frac{1}{2} g_{\mu\nu} R + \Lambda g_{\mu\nu} = 8 \pi G T_{\mu\nu}.
\label{Einstein}
\end{equation}

\noindent
We adopt the FLRW metric \cite{Friedmann,Lemaitre,Robertson,Walker}
which reflects homogeneity and isotropy

\begin{equation}
ds^2 = dt^2 - a(t)^2 \left[dr^2 + r^2 (d\theta^2 + {\rm sin}^2 d\phi^2)  \right],
\label{FLRW}
\end{equation}

\noindent
and substituting Eq. (\ref{FLRW}) into Eq.(\ref{Einstein}) gives {\it inter alia} the
Friedmann expansion equation\cite{Friedmann}

\begin{equation}
\left( \frac{\dot{a}}{a} \right)^2 = \frac{8 \pi G}{3} \rho_{TOT}
\label{Friedmann}
\end{equation}

\noindent
in which the total density $\rho_{TOT}$ has three important components

\begin{equation}
\rho_{TOT} = \rho_m + \rho_{\gamma} + \rho_{\Lambda}
\label{density}
\end{equation}

\noindent
corresponding to matter, radiation and dark energy respectively.

\bigskip

\noindent
Can the visible universe be regarded a thermodynamically isolated system? The answer is
a categoric yes. No heat ever enters or leaves  and it can be considered as if its surface
were covered by a perfect thermal insulator. It contains a {\it very} large number of particles,
much bigger than the number of molecules, $6.023 \times 10^{23}$, per mole of ideal gas.

\bigskip

\noindent
We assume that the universe is finite and that the visible universe (VU) is a 2-sphere
characterised by one scale, $R_{VU}(t)$, which is its co-moving radius, monotonically increasing
as if from $R(t=0) = 0$ at the Big Bang to $R_{VU}(t_0)$ the present time $t_0 \sim 13.8$Gy when
\begin{equation}
R_{VU} (t_0=13.8Gy) \sim 44 {\rm Gly}
\label{Rt0}
\end{equation}

\noindent
is a reasonable value. There is a subtlety after the expansion starts 
accelerating at time $t=t_{DE} \sim 9.8$Gy when the dark energy becomes
dominant. The universe then acquires an extroverse, a 2-sphere of radius $R_{ext}(t)$
which is larger than
the visible universe for $t > t_{DE}$; the present values are
\begin{equation}
R_{ext}(t_0) \sim 52 {\rm Gly} > R_{VU}(t_0) \sim 44 {\rm Gly}
\label{Rext}
\end{equation}

\noindent
This implies that the present extroverse has a volume 65\% larger than
the visible universe so that if, say, the VU contains 100 billion
galaxies, a further 65 billion galaxies have already exited the VU
but it is reasonable to assume that, despite their lack of observability,
those extra galaxies have similar dark matter and supermassive 
black holes to the ones of the visible universe.
The accelerated expansion is thus unimportant to our analysis
of the dark matter which was formed in the first three years
when $0 \leq t \leq 3y$.

\bigskip

\noindent
To set the scene, let us make an inventory of entropies of the known objects inside
the visible universe, excluding dark matter and dark energy.\\

\bigskip

\noindent
We use dimensionless entropy $S/k$.

\bigskip
\begin{center}
\noindent
Luminous matter (Baryons)  $\sim10^{80}$

\noindent
Cosmic Microwave Background   $\sim 10^{88}$

\noindent
Relic neutrinos  $\sim 10^{88}$

\noindent
Supermassive black holes  $\sim 10^{103}$
\end{center}

\noindent
For the supermassive black holes, we have made our estimate based
on a number $10^{11}$ of galaxies in the visible universe, an average SMBH mass
of $10^7 M_{\odot}$, and the black hole entropy formula

\begin{equation}
S_{BH}(M=\eta M_{\odot}) ~~ \sim ~~ 10^{78} \eta^2
\label{BHentropy}
\end{equation}

\noindent
These entropies reveal the staggering fact that of
all the present entropy in the visible universe a fraction $(1-10^{-15})$
is contained in the SMBHs. Equally striking
is that, from Eq.(\ref{BHentropy}), one SMBH with mass $10^7 M_{\odot}$
contains entropy $10^{92}$ which is $10,000$ times the entropy of
all the CMB permeating the entire visible universe.

\bigskip

\noindent
This overwhelming domination of entropy by black holes reflects the fact
that by far the most efficient way of concentrating entropy is by black holes.
This supports the idea that, for example, the dark matter inside clusters of
galaxies, imaged by gravitational lensing, exists because
in the early universe Nature created \cite{NP2020} 
very large numbers of primordial black holes
to satisfy the second law of thermodynamics.

\bigskip

\noindent
Although the PBH formation in the early universe is a dynamical question
which, to be rigorous, would require solution of Boltzmann's transport
equation, Eq.(\ref{transport}), for $\sim 10^{80}$ particles which is
far beyond the capability of any computer, nevertheless we can
specify the eras during which the PBHs were formed. The mass of
a PBH is determined by the horizon size to be
\begin{equation}
M_{PBH} \simeq 10^5 M_{\odot} \left( \frac {t}{ 1 {\rm sec}} \right)
\label{PBHMASS}
\end{equation}

\bigskip

\noindent
The MACHO Collaboration\cite{Alcock} found that MACHOs up to
mass $25M_{\odot}$ could account for no more that $10\%$ (or $20\%$
at a stretch) of the dark matter. We shall assume that the MACHOs
they did discover included PBHs. In any case, the other $90\%$
of the dark matter PBHs must be heavier than $25M_{\odot}$
which, according to Eq. (\ref{PBHMASS}), requires

\begin{equation}
t \geq t_1 \equiv 0.00025 {\rm sec}
\label{t1}
\end{equation}

\bigskip

\noindent
For MACHO searched by a microlensing experiment which targets
the Magellanic clouds, the duration ($\tau$) of the light curves for full-width
at half maximum is given by

\begin{equation}
\tau \simeq 0.2 {\rm  year} \left( \frac{M_{MACHO}}{M_{\odot}} \right)^{\frac{1}{2}}
\label{tau}
\end{equation}

\bigskip

\noindent
and so if we find it impracticable to wait for more than $10 yrs$ to measure a light 
curve we require $M_{PBH} \leq 2500 M_{\odot}$ and according to
Eq.(\ref{PBHMASS}) such a PBH is formed before $t_2$ given by

\begin{equation}
t_2 \equiv 0.025 {\rm sec}
\label{t2}
\end{equation}

\bigskip

\noindent
Let us define PIMBHs as residing in the mass range

\begin{equation}
25 M_{\odot} \leq M_{PIMBH} \leq 10^6 M
\label{MassPIMBH}
\end{equation}

\bigskip

\noindent
then PIMBHs were formed before a time $t_3$ which is

\begin{equation}
t_3 = 10 {\rm sec}
\label{t3}
\end{equation}

\noindent
Let us define supermassive black holes (SMBHs) to be those black holes
in the mass range

\begin{equation}
10^6 M_{\odot} \leq M_{SMBH} \leq 10^{12}M_{\odot}
\label{MassSMBH}
\end{equation}

\bigskip

\noindent
so that SMPBHs are formed\cite{PHF999} before $t_4$ given by

\begin{equation}
t_4 = 10^7 {\rm seconds} \simeq ~ 3 {\rm years}
\label{t4}
\end{equation}

\bigskip

\noindent
To summarise, all of the cosmological dark matter and the supermassive
black holes are formed during the first three years.
Normalising the FLRW scale factor $a(t)$ in Eq.(\ref{FLRW}) by
$a(t_o) \equiv 1$ at the present time the values of $a(t)$ and the 
corresponding values for the radius $R_{VU}(t)$ of
the visible universe are:

\bigskip

\noindent
$t_1 = 0.00025s$ ~~~ $a(t_1)= 2.7 \times 10^{-12}$ ~~~ $R_{VU}(t_1) = 0.14$ ly.

\bigskip

\noindent
$t_2 = 0.025s$ ~~~~~~ $a(t_2)= 2.7 \times 10^{-11}$ ~~~ $R_{VU}(t_2) = 1.4$ ly.

\bigskip

\noindent
$t_3 = 10 s$ ~~~~~~~~~ $a(t_3)= 5.4 \times 10^{-10}$ ~~~ $R_{VU}(t_3) =  28$ ly.

\bigskip

\noindent
$t_4 = 10^7s$ ~~~~~~ ~~$a(t_4)= 5.4 \times 10^{-7}$ ~~~~ $R_{VU}(t_1) = 28$ kly.

\bigskip

\noindent
The intermediate-mass PIMBHs are formed between cosmic times $t_1$ and $t_3$, 
with the most readily detectable made between $t_1$ and $t_2$. The supermassive black holes
then appeared between times $t_3 \sim 10s$ and $t_4 \sim 3y$. All of the dark matter was formed
during the first three years after the Big Bang, just as all the luminous matter
was formed during the first three minutes.

\newpage

\section{On the Nature of Dark Matter}

\bigskip

\noindent
In this Section, we shall first discuss WIMPs (subsection 2.1) and axions (subsection 2.2) which are the two most likely
candidates, in that order, for the constituents of cosmological dark matter which arise from extensions
of the standard model of particle phenomenology. We shall
then return to primordial black holes (subsection 2.3) whose motivation is based instead on 
entropy as discussed in Section 1.

\bigskip

\subsection{WIMPs} 

\noindent
By Weakly Interacting Massive Particle (WIMP) is generally meant an unidentified elementary particle with mass in the range between 10 GeV and 1000 GeV and with scattering cross section with nucleons (N) satisfying, according to the latest unsuccessful WIMP direct searches,

\begin{equation}
\sigma_{WIMP-N} < 10^{-46} {\rm cm}^2 
\label{WIMPsigma}
\end{equation}

\noindent
which is roughly comparable to the characteristic 
strength of the known weak interaction. 

\bigskip

\noindent
The WIMP particle must be electrically neutral and be stable or have an extremely long lifetime, longer
than the age of the universe. In model-building, this stability may be achieved by an {\it ad hoc} 
discrete symmetry, for example a $Z_2$ symmetry under which all the standard model particles are even and others are odd. If the discrete symmetry is unbroken, the lightest odd state must be stable and therefore a candidate for a dark matter. 

\bigskip

\noindent
By far the most popular WIMP example comes from electroweak supersymmetry where a discrete R symmetry has the value R=+1 for the standard model particles and R=-1 for all the sparticles. Such an R parity is less {\it ad hoc}, being essential to prevent too-fast proton decay. The lightest R=-1 particle is stable and, if not a gravitino which has the problem of too-slow decay in the early universe, it is the neutralino, 
a linear combination of zino, bino and higgsino. The neutralino provides an attractive dark matter
candidate.

\bigskip

\noindent
It is worth briefly recalling the history of electroweak supersymmetry. The standard model 
\cite{Glashow,Weinberg,GIM,Hooft10} was in place by 1971 and its biggest theoretical problem was that, unlike QED with only logarithmic divergences, the scalar sector of the standard model generates quadratic divergences which, unless cancelled within a quiver-type construction \cite{PFquiver}, destabilise the mass of the BEH boson. When supersymmetric field theories were invented\cite{Ramond,WessZumino} in 1974, they provided a mathematical solution of the quadratic divergence problem and immediately became popular. Even more so in 1983 when the neutralino was identified\cite{Goldberg} as a dark matter candidate and more so again in 1991 when it was pointed out\cite{ADF} that grand unification apparently works better with hypothetical supersymmetric partners included.

\bigskip

\subsubsection{Direct detection}

\noindent
With all of this support for supersymmetry, it is natural to take seriously the dark matter
candidate which such a theory predicts. It is a WIMP with mass typically 10-1000GeV
and experiencing weak interactions which would suggest a detectable scattering
cross-section from nuclei in {\it direct} detection experiments.

\bigskip

\noindent
Some of the detectors for WIMPs have been built using liquid xenon\cite{XENON1,XENON2,XENON3} .These have 
produced the strongest upper limits on the existence of WIMPs such as
the cross-section quoted above in Eq.(\ref{WIMPsigma}). 

\bigskip

\noindent
When a WIMP passes through a detector, it can interact with a nucleus which will recoil.
The idea is to detect the small energy which is transferred. Experiments may have
$1,000$ kgs up to $10,000$ kgs of detector. Such an experiment needs knowledge
of the WIMP-nucleus cross-section and the distribution of WIMPs in the galactic halo.

\bigskip

\noindent
The WIMP-nucleus interaction can be spin-independent (SI) or spin-dependent (SD).
For SI, the nucleons add to a $A^2$ coherence factor. If the WIMP is heavier than the
nucleus this becomes $A^4$, so heavy nuclei are the best targets. This is why germanium
($A=74$) is a popular choice because $74^4 \simeq 3\times 10^7$.
For SD, the WIMP interacts only with the total spin of the nucleus, and the factor $A^2$
is lost. It needs a nuclus with a nonzero spin. In the case of the CDMS experiment,
for example,
it uses\cite{CDMS1,CDMS2,CDMS3} a mixture of Ge-74 (spinless) and Ge-73 (spin$ = \frac{9}{2}$)
in order to be sensitive to both SI and SD scattering of WIMPs.

\bigskip

\noindent
Concerning the astrophysics, the count rates are highest for slow WIMPs and
go to zero for WIMP velocity $540 km/s$ which is the escape velocity from the
halo. One needs also to consider the density profile of the halo which is assumed
to follow the NFW profile\cite{NFW1,NFW2} obtained from numerical simulations. This peaks at the galactic 
centre and the density falls, $\rho \sim r^{-2}$, for large $r$. This density may be lumpy 
because the Milky Way was formed in part by mergers. The Sun is $24,000 ly$ from
$Sag A*$ and is moving at $\sim 250 km/s$ around the core. The average density
of WIMPs is $0.4 GeV/cm^3$ and their relative velocity is what determines the flux.
An annual modulation is expected due to the Earth's orbit around the Sun. 

\bigskip

\noindent
Background noise for direct detection can arise from cosmic rays and solar flares
so the detectors are placed deep underground. For example, the Homestake
Mine in South Dakota is almost one mile deep. Nevertheless, even there
radioactivity of the rock, due to {\it e.g.} naturally occurring radon, must be
taken into account.

\bigskip

\noindent
A large detector for WIMPs, the LZ Dark Matter Experiment using 
seven tons of liquid xenon, is planned for SURF (Sanford Underground Research Facility) 
in South Dakota\cite{LZ}. 

\subsubsection{Production}

\noindent
Another way of finding the WIMPs is to look for {\it production} in pairs at
a particle collider such as the Large Hadron Collider (LHC) which is presently
the highest-energy machine in the world with proton-proton collisions at
13 TeV centre-of-mass energy.

\bigskip

\noindent
Dark matter itself is not detected in a production experiment like LHC. Instead
one looks for an apparent violation of energy conservation. If WIMPs were produced
at the LHC, the signature would be high-transverse-momentum jets which
are easily detected. Pair production of WIMPs
should be associated by 1, 2 or more such jets.

\bigskip

\noindent
If WIMPs are produced and detected, it will still need astrophysical evidence that
it is the dark matter.

\bigskip

\subsubsection{Indirect detection}

\bigskip

\noindent
A third method to search for WIMPs is to seek astrophysical signals of WiMP
annihilation products. Many WIMPs are their own antiparticles and annihilate
among themselves into a variety of lighter particles. The end states include
$e^+,$ $\gamma$ and $\nu$. These are sought by detectors on satellites
in space, strings of phototubes embedded in ice at the South Pole, and
other techniques.

\bigskip

\noindent
The strongest signals come from regions most abundant in WIMPs,
{\it viz} the centres of the Earth and Sun, the galactic centre and dwarf 
galaxies near the Milky Way. It needs a WIMP density high enough for
them to collide and annihilate. WIMP annihilations in the early universe
are important in order that the relic density of dark matter can agree
with that observed at the present time. 

\bigskip

\noindent
Now the average WIMP density is so low that, in general, they never collide.
The only places they will appreciably annihilate is in regions of
especially high WIMP overdensity. The Milky Way has a higher density
than the universe's average.

\bigskip

\noindent
WIMPs traveling through the Earth have a probability of about one in ten
billion to hit a nucleus and lose sufficient energy to be captured and
pulled to the Earth's centre of gravity. There they start annihilating.
The $e^+$'s cannot escape the Earth's core because of electromagnetic
forces, but the neutrino products of WIMP annihilation can. The failure
to observe these in surface neutrino detectors provides a useful upper
limit on the WIMP density at the centre of the Earth.

\bigskip

\noindent
A similar analysis applies {\it mutatis mutandis}, to the centre of the Sun.
Searches for $\nu$'s from the centres of the Earth and Sun
are continuing.

\bigskip

\noindent
We can go to larger distances from the Earth whereupon a good candidate
for WIMP concentration is the Milky Way core at a distance of 24,000 ly.
Computer simulations of the dark matter galactic profile suggest a higher
density there. The situation in the vicinity of Sag$A^*$ is not straightforward
because (i) merging of black holes into the supermassive black hole
could have knocked WIMPs away, and (ii) competing astrophysical processes
might be difficult to disentangle.

\bigskip

\noindent
Dwarf galaxies can produce cleaner signals for WIMPs. Inside the Milky Way 
are many substructures, including dwarf galaxies which are between $10^{-6}$
and $10^{-3}$ of the mass of the galaxy. There are at least twenty dwarf galaxies
inside the Milky Way with exceptionally high ratio of dark matter to luminous matter,
and these may provide the best environments for WIMPs.

\bigskip

\noindent
The intricate decay chains for the WIMP annihilations into standard model particles
can be quite complex, but can be calculated with the most certainty for the
special case of the MSSM neutralino. Generally the total mass of the final
products may add to about ten percent of the original progenitor WIMP mass.

\bigskip

\noindent
There is a worldwide search for the products of WIMP annihilation using, {\it e.g.},
satellites and the IceCube detector\cite{IceCube1,IceCube2,IceCube3}
at the South Pole. Excesses of $e^+$'s and
$\gamma$'s could provide us with the smoking gun for such indirect detection.

\bigskip

\subsection{Axions}

\bigskip

\noindent
The axion particle is now believed, 
if it exists, to lie in the mass range

\begin{equation}
10^{-6} eV < M < 10^{-3} eV 
\label{axionmass}
\end{equation}

\bigskip

\noindent
The lagrangian originally proposed for Quantum Chromodymamics 
(QCD) was of the simple form, analogous to Quantum Electrodynamics (QED),

\begin{equation}
{\cal L}_{QCD} = - \frac{1}{4} G_{\mu\nu}^{\alpha} G_{\alpha}^{\mu\nu} 
- \frac{1}{2} \Sigma_i 
\bar{q}_{i,a} \gamma^{\mu} D_{\mu}^{ab} q_{i,b}
\label{QCD}
\end{equation}

\noindent
summed over the six quark flavors.
The simplicity of Eq.(\ref{QCD}) was only temporary and became more complicated in 1975 
by the discovery of instantons which dictated that an additional term 
be allowed in the QCD lagrangian

\begin{equation}
\Delta {\cal L}_{QCD} = \frac{\Theta}{64 \pi^2} G_{\mu\nu}^{\alpha} \tilde{G}_{\alpha}^{\mu\nu}
\label{GGdual}
\end{equation}

\noindent
must be added where $\tilde{G}_{\mu\nu}$ is the dual of $G_{\mu\nu}$. Although this 
extra term is an exact derivative, it cannot be discarded as a surface term because 
there is now a topologically nontrivial QCD vacuum
with an infinite number of different values of the spacetime integral over Eq.(4) 
all of which correspond to $G_{\mu\nu} = 0$. Normalized as in Eq.(4), the 
spacetime integral of this term must be an integer, and an instanton configuation 
changes this integer, or Pontryagin number, by unity.

\bigskip

\noindent
When the quark masses are complex, an instanton changes not only $\Theta$
but also the phase of the quark mass matrix ${\cal M}_{quark}$ and the full phase 
to be considered is

\begin{equation}
\bar{\Theta}  = \Theta + arg det ||{\cal M}_{quark}|| 
\label{Thetabar}
\end{equation}

\noindent
The additional term, Eq.(4), violates P and CP, and contributes to the neutron electric
dipole moment whose upper limit provides a constraint
\begin{equation}
\bar{\Theta}  < 10^{-9} 
\label{ThetabarLimit}
\end{equation}

\noindent
which fine-tuning is the strong CP problem.

\bigskip

\noindent
The hypothetical axion particle then arises from a method to resolve Eq.(6),  
based on the Peccei-Quinn mechanism which introduces a new global $U(1)_{PQ}$
symmetry which allows the vacuum to relax to $\bar{\Theta} = 0$. Because this 
$U(1)_{PQ}$ symmetry is spontaneously broken, it gives rise to a light 
pseudoscalar axion\cite{Weinberg2,Wilczek} with mass in the range $100keV < M < 1MeV$. 

\bigskip

\noindent
An axion in this mass range was excluded experimentally but then the theory 
was modified\cite{DFS,Kim,Zhitnitsky,SVZ} to one with an invisible axion where the $U(1)_{PQ}$
symmetry is broken at a much higher scale $f_a$ and the coupling of the 
axion correspondingly suppressed. Nevertheless, experiments to detect 
such so-called invisible axions were proposed, firstly using resonant
microwave cavities then using other techniques discussed in
sections 2.2.2 and 2.2.3.

\bigskip

\subsubsection{Resonant Microwave Cavities}

\bigskip

\noindent
The first method to detect the invisible axions was suggested by
Sikivie\cite{Sikivie} in 1983. The idea is that the dark matter axions
will move through a microwave cavity in a strong magnetic field
and be resonantly converted to photons. The very weak coupling 
of the axion is compensated by their large number, typically
$\sim 10^{14} cm^{-3}$ if axions form all the dark matter.

\bigskip

\noindent
The microwave signal will be almost monochromatic at a frequency
corresponding to the axion mass, broadened upward because of the
axion's virial distribution. The expected velocity is $\sim 10^{-3} c$
which leads to a spread in energy $\delta E/E \sim 10^{-6}$.

\bigskip

\noindent
The lagrangian for the axion coupling is

\begin{equation}
{\cal L} = \left( \frac{\alpha g_{a\gamma\gamma}}{2 \pi f_a} \right)
a {\bf E}.{\bf B}.
\label{aEB}
\end{equation}

\bigskip

\noindent
The resonant modes which couple to axions are transverse magnetic 
(TM) modes. The predicted power from axion $\longrightarrow$ photon 
conversion is\cite{Sikivie2}

\begin{equation}
{\cal P}_a = \left( \frac{\alpha g_{a\gamma\gamma}}{2 \pi f_a} \right)^2 V B^2 \rho_A C
M_A^{-1} {\rm min}(Q_L. Q_a)
\label{Pcavity}
\end{equation}

\noindent 
where v is the cavity volume, B the magnetic field, $\rho_a$ is the axion volume, $C$ is
a form factor characterising overlap of a specific TM mode, $Q_L$ is the loaded quality
factor and $Q_a$ is  the axion quality factor.

\bigskip

\noindent
The mass range for dark matter axions centres around $1 \mu eV$ to $1 meV$, although
extensions of this mass range in both directions are being studied. The converted
photon frequencies are in the range MHz to THz. Experiments have been designed to look
at the lower frequencies in this range but to maintain a high quality factor only a few kHz
can be scanned at a time.

\bigskip

\noindent
The scan rate is determined by the time it takes a possible axion signal to be above 
the cavity's intrinsic noise, according to the radiometer equation

\begin{equation}
SNR = \frac{{\cal P}_a}{\bar{{\cal P}}_N} \sqrt{Bt} = \frac{{\cal P}_a}{k T} \sqrt{\frac{t}{B}}
\label{radiometer}
\end{equation}

\noindent
with ${\cal P}_a$ the power generated by axion-photon conversion, 
${\cal P}_N = kBT$ is the cavity noise power, $B$ is the signal
bandwidth, $t$ is the integration time, $k$ is Boltzmann's constant
and $T$ is the temperature.

\bigskip

\noindent
From Eq.(\ref{radiometer}), a tiny signal power can be amplified
by increasing ${\cal P}_a \propto VB^2$, increasing the time $t$ 
or minimising $T$. Most of the research is directed to lowering
 the intrinsic noise. The earliest such experiments were carried 
 out at Brookhaven National Laboratory\cite{Panfilis}, then at
 the University of Florida\cite{Hagmann}.
 
 \bigskip
 
 \noindent
 The ADMX (Axion Dark Matter eXperiment) at LLNL is one
 example \cite{ADMX} of a second-generation experiment. ADMX uses a
 $8.5$ Tesla superconducting magnet 110cm in length and
 a 200 liter stainless steel microwave cavity plated with ultra-pure
 copper. An adjustable antenna is put through the top cavity plate
 and its signal is boosted by extremely low noise cryogenic
 amplifiers which are the most important limiting factor on
 axion sensitivity.
 
 \bigskip
 
 \noindent
 More recent cavity designs are discussed in \cite{Stern}.

\bigskip

\subsubsection{Axion Helioscopes}

\bigskip

\noindent
Axions produced in the Solar core free-stream out to be detected
on Earth when they convert into low-energy X-rays as they pass through
a strong magnetic field. The flux of axions produced in the Sun should
have a thermal spectrum with a mean energy of a few keV, and the
integrated flux art the Earth is expected\cite{Zioutas} to be $\sim 10^{11} cm^{-2}s^{-1}$.

\bigskip

\noindent
Consider a solar axion passing through a magnetic field {\bf B} with length $L$
then its probability $P$ of conversion into a photon is given by \cite{Sikivie2}

\begin{equation}
P = \left( \frac{\alpha g_{a\gamma\gamma} BL}{4 \pi f_a} \right)^2
2L^2 \frac{1 - {\rm cos}(qL)}{(qL)^2}
\label{probability}
\end{equation}

\noindent
in which $g_{a\gamma\gamma}$ is the axion-photon coupling and $q$ is the 
axion-photon momentum difference given by $q=m_a^2/2E$ where $E$ is
the photon energy. Maximum conversion of axions to photons occurs when 
their fields stay in phase over the length $L$ of the magnet. This requires\cite{Bibber}
that $qL < \pi$. When the axion mass is small, $q \longrightarrow 0$ and
the axion$\rightarrow$photon conversion is greatest. More massive axions 
tend to go out of phase, but there is a method to compensate by adding a buffer
gas which imparts an effective mass to the photon. Different axion masses
can then be tuned by varying the gas pressure.

\bigskip

\noindent
The first axion helioscope was built\cite{Lazarus} at BNL (Brookhaven) in 1992 but the limits
it obtained were far outside the expected parameters of invisible axion theories.
Follow up experiments at the University of Tokyo\cite{Inoue} in 2007-08 obtained more
stringent limits such as $g_{a\gamma\gamma}<(5.6-13.4) \times 10^{-10} GeV^{-1}$
for axions on the mass range $0.84 eV < m_a < 1 eV$.

\bigskip

\noindent
Upgraded experiments have been constructed at CERN and the 
University of Tokyo. At CERN, the experiment \cite{Andriamonje} is called CAST
(= CERN Axion Solar Telescope). CAST uses an LHC magnet
of length $L=9.3m$ and magnetic field $B=9$ Tesla. It tracks the 
Sun for 90 minutes a day using a rail system and the double magnet bore
permits four X-ray detectors, one at each end of each bore.
CAST has achieved a limit of 
$g_{a\gamma\gamma}/f_a < 7.6 \times 10^{-8} GeV^{-1}$.
$^3He$ and $^4He$ buffer gases are being used to
extend the searched region of axion mass.

\bigskip

\noindent
More sensitive than CAST is TASTE (=Troisk Axion Solar Telescope Experiment)
\cite{Anastassopoulos}.

\subsubsection{Laser Methods}

\bigskip

\noindent
One might expect that the unique coherent states of photons in lasers can
form a detection method for axions, and indeed we shall discuss two of the
possibilities for scattering off the laser photons, designated $\gamma_{LAS}$.

\bigskip

\noindent
There will be a transverse magnetic field which itself creates virtual photons,
designated $\gamma^*$. The idea then is to create axions by the scattering process

\begin{equation}
\gamma_{LAS} + \gamma^* \longrightarrow a
\label{laser}
\end{equation}

\bigskip

\noindent
We may look for disappearance of polarised laser photons as they convert
into axions by a magneto-optical effect of the vacuum. This arises from a term

\begin{equation}
a {\bf E.}{\bf B}
\label{EB}
\end{equation}

\noindent
which is an anomalous coupling.

\bigskip

\noindent
Such an axion search has been carried out by a Rochester-Brookhaven-
Fermilab-Trieste (RBFT) group \cite{Cameron1991} which at one time found a preliminary
positive signal \cite{Zavattini} that led to theories of vacuum dichroism
in which the two circular polarisations are differentiated. The polarisation
of the laser beam can be examined. The original suggestion of a
discovery was not supported by further data, but it has inspired a
strong group of searchers for axion-like-particles at the DESY
Laboratory.

\bigskip

\noindent
As a second example for axion searches using lasers
we briefly discuss what has been called, dramatically, "light shining through walls" as
suggested in \cite{Bibber2}. Polarised laser photons pass through
a magnetic field with ${\bf E} || {\bf B}$ and axions when produced
pass through an absorber (the wall) and are reconverted to axions
on the other side \cite{Adler}.

\bigskip

\noindent
The probability for a photon to convert into an
axion in the axion-source region is

\begin{equation}
{\cal P}_{\gamma \rightarrow a} \propto
\frac{1}{4} \left( \frac{\alpha g_{a\gamma\gamma}}{2 \pi f_a} BL \right)^2
\frac{1-{\rm cos}(qL)}{(qL)^2}
\label{convert}
\end{equation}

\noindent
and the probability for the axion to reconvert to an observable photon is 
the same as Eq.(\ref{convert}) and so the total probability for
detection of photon$\longrightarrow$axion$\longrightarrow$photon is\cite{Battesti}

\begin{equation}
{\cal P}_{\gamma \rightarrow a \rightarrow \gamma} = {\cal P}_{\gamma \rightarrow a}^2
\label{reconvert}
\end{equation}

\bigskip

\noindent
There is a maximum detectable axion mass because of the
oscillation length becoming shorter than the magnetic
field length. The RBFT group. already mentioned, found \cite{Cameron2}
an upper limit on the axion-photon coupling of
$g_{a\gamma\gamma} < 6.7 \times 10^{-7} {\rm GeV}^{-1}$
for axions with mass $m_a < 1 meV$.

\bigskip

\noindent
It has been shown that this "light shining though walls"
experiment can be resonantly enhanced \cite{Sikivie3} by encompassing both
the production and reconversion magnets in matched
Fabry-Perot optical resonators.

\bigskip

\noindent
Laser induced fluorescence in rare-earth doped materials is being pursued\cite{Braggio}.

\bigskip

\subsection{Primordial black holes}

\bigskip

\noindent
The idea that primordial black holes (PBHs) might be formed in the early universe
was first proposed \cite{NovikovZeldovich1967} by Novikov and Zeldovich
in the Soviet Union
in 1967. In 1974, the same idea occurred independently in the West
to Carr and Hawking\cite{CarrHawking}.
A year later in 1975 Chapline \cite{Chapline}
was the first to suggest that the dark matter could be made from PBHs,
$DM=PBHs$, which was a prescient idea. At that time, PBHs were believed to be 
orders of magnitude lighter than the Sun, the
most popular particle theory dark matter candidates, WIMPs and axions,
had not yet been invented, and microlensing experiments were unknown.

\bigskip

\noindent
Forty years later, in 2015, we proposed in Frampton \cite{NP2020} instead the idea, $DM=PIMBHs$,
where the $PIMBHs$ are many times the mass of the Sun in the intermediate-mass(IM)
region between stellar mass and supermassive black holes. At the time, we were unaware 
of Chapline's work.

\bigskip

\noindent
Important and influential were the data obtained by the MACHO Collaboration\cite{Alcock}
including examples of microlensing light curves for lens masses up to
almost $25M_{\odot}$; it is entirely possible that if that experiment
had continued beyond 1999, the dark matter PIMBHs could have been 
discovered although the possible additional time was limited because the
Mount Stromlo Observatory was destroyed in the Canberra bushfire of 2003. 

\bigskip

\noindent
There were also the inventions of the WIMP and axion particles where the former
became by far the most popular candidate for dark matter. One realization
expressed in the 2015 paper \cite{NP2020} was that the failure of the LHC to confirm
the presence of electroweak supersymmetry at the same time weakened the
motivation for the WIMP and therefore made more likely an astrophysical
solution. It was mentioned that the most promising test
was by microlensing and this was stressed further in a joint paper, Chapline
and Frampton \cite{ChaplineFrampton} in 2016.

\bigskip

\noindent
Several other senior physicists have thought about entropy of the universe
or about black holes as dark matter. Six examples are 
Carr \cite{CKSY,CKS},  
Garcia-Bellido\cite{GB1,GarciaBellido2,GarciaBellido3,GarciaBellido4,GarciaBellido5,GarciaBellido6,
GarciaBellido7,GarciaBellido8,GarciaBellido10,GarciaBellido11}, Jacobson\cite{Jacobson},
Linde\cite{Linde1,Linde2,Linde3,Linde4,Linde5}, 
Rees\cite{Rees1,Rees2,Rees3}
and Verlinde \cite{EV1,EV2}.  
We apologise to any author not mentioned.

\bigskip

\noindent
Massive Compact Halo Objects (MACHOs) are commonly defined by the notion 
of compact objects used in astrophysics as the end products of stellar evolution 
when the nuclear fuel has been expended. They are usually defined to include white 
dwarfs, neutron stars, black holes, brown dwarfs and unassociated planets, all equally 
hard to detect because none of them emit significant electromagnetic radiation.

\bigskip

\noindent
This narrow definition implies, however, that MACHOs are composed of baryonic matter 
which is too restrictive in the special case of black holes. It is here posited that black holes 
of arbitrarily high mass up to $10^{12} M_{\odot}$ can be produced primordially. Nevertheless the acronym MACHO still nicely applies to dark matter 
PIMBHs which are massive, compact, and in the halo.

\bigskip

\noindent
Unlike the axion and WIMP elementary particles invented within the framework
of quantum field theory which would have a definite mass, the black holes 
arising as classical solutions of Einstein's equations have a range of masses. 
The lightest PBH which has survived for the age of the universe has a lower mass limit
$M_{PBH} >10^{-18} M_{\odot} \simeq 10^{36}$ TeV, thirty-six orders of magnitude 
heavier than the WIMP with $10 {\rm GeV} \leq M_{WIMP} \leq 1,000$ GeV. This
lower limit on the PBH mass comes from the lifetime formula derivable from 
Hawking radiation\cite{Hawking1}

\bigskip

\noindent
Because of observational constraints, the dark matter constituents must generally be another twenty orders of magnitude more massive than the lower limit in Eq.(9). We assert that most dark matter black holes are in the mass range between twenty-five and a trillion times the solar mass. The designation
intermediate-mass for PIMBHs is appropriate for 
$25 M_{\odot} \leq H_{PIMBH} \leq 10^6 M_{\odot}$ because they lie in mass above 
stellar-mass black holes with $M_{BH} \leq 25 M_{\odot}$ and below the supermassive 
black holes with $M_{BH} \geq 10^6 M_{\odot}$ which reside in galactic cores.

\bigskip

\noindent
We shall discuss three methods which can be used to search 
for PIMBHs: Wide binaries, CMB distortion and Microlensing
in subsections 2.3.1, 2.3.2 and 2.3.3 respectively. 

\bigskip

\subsubsection{Wide binaries}

There exist in the Milky Way pairs of stars which are gravitationally 
bound binaries with a separation more than 0.1pc. These wide binaries retain their original orbital parameters unless compelled to change them by gravitational influences, for example, due to nearby PIMBHs.
Because of their very low binding energy, wide binaries are particularly sensitive to 
gravitational perturbations and can be used to place an upper limit on, or to detect, 
PIMBHs.

\bigskip
  
\noindent
The history of employing this ingenious technique is regretfully checkered. 
In 2004 a fatally strong constraint was claimed by an Ohio State University 
group \cite{Gould} in a paper entitled {\it End of the MACHO Era},
where stellar and higher mass constituents of dark matter were 
totally excluded.

\bigskip

\noindent
Five years later in 2009, however, another group this time from 
Cambridge University \cite{QuinnEtAl} reanalyzed the available data on wide binaries 
and reached an opposite conclusion. They questioned whether any rigorous 
constraint on MACHOs could yet be claimed, especially as one of the 
important binaries in the earlier sample had been misidentified.

\bigskip

\noindent
In 2014, the most recent publication on wide binaries appeared\cite{Monroy}
which claims that, after all, some bound on MACHOs can be claimed,
so this approach is still very much alive.

\bigskip

\noindent
Because of the checkered history, however, it seems wisest to proceed 
with caution in reaching any categoric conclusions from wide binaries,
but to acknowledge that they represent a potentially useful source 
both of constraints on, and the possible discovery of, dark matter PIMBHs.

\bigskip

\subsubsection{Distortion of the CMB}

\bigskip

\noindent
This approach hinges on the phenomenon of accretion of gas onto the PIMBHs. 
The X-rays emitted by such accretion of gas are downgraded in frequency 
by cosmic expansion and by Thomson scattering becoming microwaves which 
distort the CMB, especially with regard to its very-precisely-measured
black-body spectrum.

\bigskip

\noindent
One early and detailed calculation of this effect by Ricotti, Ostriker and Mack (ROM)
\cite{ROM} has been very influential. ROM employed a specific model for the 
accretion, the Bondi model, and carried
through the computation all the way up to the point of comparison with data 
from FIRAS on CMB spectral distortions, where FIRAS was a detector
attached to the COBE satellite. ROM concluded that MACHOs with many solar
masses could provide no more than a tiny fraction $\sim 10^{-4}$ of the
dark matter.

\bigskip

\noindent
The implication of ROM was that dark matter constituents with many solar masses
were excluded unless the ROM calculation was in error by four orders of
magnitude. Surprisingly the latter was the correct conclusion, as confirmed
in 2016 by Ostriker\cite{JPO}. There were grounds for suspecting this to be the
case from observations of the X-rays from the supermassive black hole
at the centre of galaxy M87 which were at least four orders of magnitude
below the prediction by the Bondi model that assumes spherical
symmetry and radial inflow which are questionable assumptions.

\bigskip

\noindent
More recent papers have made exclusion plots of the allowed
fraction of dark matter versus MACHO mass but their limits are
sometimes far too severe for the same reason as for ROM. To mention
one well-known analysis, Ali-Ha\"imoud and Kamionkowski\cite{AliKam}
obtain upper limits which, while somewhat softer than ROM, remain 
suspiciously strong because, like ROM, they employ quasi-spherical accretion.

\bigskip

\noindent
A recent claim by Zumalacarregui and Seljak\cite{SelZum}
that absence of lensing by supernovae provides a
contradiction to dark matter comprised of many-solar-mass
constituents has been criticised in a reanalysis
by Garcia-Bellido, Clesse and Fleury\cite{GCF}.

\bigskip

\subsubsection{Microlensing}

\noindent
Microlensing is the most direct experimental method and has the big advantage 
that it has successfully found examples of MACHOs. The MACHO Collaboration 
used a method which had been proposed by Paczynski \cite{Paczynski} where the 
amplification of a distant source by an intermediate gravitational lens 
is observed. Unbeknownst to Paczynski, the microlensing equations had
been calculated much earlier by Einstein \cite{Einstein3}
who did not publish because he thought such measurements were
impracticable. However, Einstein was overly pessimistic
because the MACHO Collaboration discovered several striking 
microlensing light curves. 

\bigskip

\noindent
The method certainly worked well\cite{Alcock} for $M < 25 M_{\odot}$ 
and so should work equally well for $M > 25 M_{\odot}$
provided one can devise a suitable algorithm and computer 
program to scan enough sources.
The longevity of a given lensing event is proportional to the 
square root of the lensing mass and numerically is given by ($t$ is longevity)
\begin{equation}
t \simeq 0.2 {\rm yr} \left( \frac{M_{MACHO}}{1 M_{\odot}} \right)^{\frac{1}{2}}
\label{timelens}
\end{equation}

\noindent
where a transit velocity $200$km/s is assumed for the lensing object.

\bigskip

\noindent
The MACHO Collaboration investigated lensing events with longevities 
ranging between about two hours and one year. From Eq.(\ref{timelens}) 
this corresponds to MACHO masses between approximately $10^{-6}M_{\odot}$
and $25 M_{\odot}$.

\bigskip

\noindent
The total number and masses of objects discovered by the MACHO Collaboration 
could not account for all the dark matter known to exist in the Milky Way. At most 
$10\%$ could be explained.

\bigskip

\noindent
What is being suggested is that the other 90\% of the dark matter in the Milky Way
is in the form of MACHOs which are more massive than those detected by the 
MACHO Collaboration, and which almost certainly could be detected by a straightforward 
extension of their techniques. In particular, the expected microlensing events have a 
duration ranging from one year ($25 M_{\odot}$) to ten years ($2,500 M_{\odot}$), which
is the practical limit for a feasible experiment.

\bigskip

\begin{table}
\caption{Microlensing duration $\hat{t}$ for the case of $n$ PIMBHs per halo
with PIMBH mass = $\eta M_{\odot}$, halo mass = $10^{12} M_{\odot}$ and universe mass = $10^{23}M_{\odot}$}
\begin{center}
\begin{tabular}{||c|c|c|c||c||}
\hline
n / Halo & M=$\eta M_{\odot}$  & Halo Entropy & Universe Entropy & Duration   \\
  & $\eta$ & $ (S_{halo}/k)$ & $(S_U/k)$  & $\hat{t}$ (years) \\
\hline
\hline
$4 \times 10^{10}$ & 25 & $2.5 \times 10^{90}$ & $2.5 \times 10^{101}$ &  1  \\
\hline
$10^{10}$  & 100 & $10^{91}$ & $10^{102}$ & 2 \\
\hline
$4 \times 10^8$ & 2500 & $2.5 \times 10^{92}$ & $2.5 \times 10^{103}$ & 10 \\
\hline
\hline
\end{tabular}
\end{center}
\label{longevity}
\end{table}

\noindent
We have simplified the visible universe, without losing anything important by regarding 
it as containing exactly $10^{11}$ galaxies, each with mass 
(dominantly dark matter) of exactly $10^{12} M_{\odot}$. 
The first three columns of Table 1 consider one halo of dark matter. To a first approximation, we can temporarily ignore the normal matter. The fourth column gives the additive entropy of the universe for well separated halos and the fifth column gives the corresponding microlensing event longevity in years.
For a black hole with mass $M_{BH} = \eta M_{\odot}$, the dimensionless entropy is 
$S_{BH}/k \sim 10^{77} \eta^2$. 

\bigskip

\noindent
We note that the entries in the
fourth column of Table 1 are of the same order
of magnitude as the value $S_{SMBH}/k \sim 10^{103}$ quoted in Section 1
for supermassive black holes. In a study \cite{FL} made in 2009 entropies of the
universe as high as $S_U/k \sim 10^{106}$ ( a million googols)
were found for PIMBHs with mass
$M_{PIMBH} = 10^5 M_{\odot}$. According to Eq.(\ref{timelens}) the microlensing
light curves would then last several decades which seems impracticable
except that a more sophisticated data analysis, in the future, might permit identification
using only a fraction of the light curve.
 
\bigskip

\noindent
For a given total halo mass, 
$M_{Halo} = 10^{12}M_{\odot}$, a smaller number of heavier black holes gives 
higher entropy because $S_{BH}  \propto M_{BH}^2$ .  Such arguments 
using the concept of the entropy of the universe \cite{FHKR} have for long
been suggestive of more black holes than the stellar and supermassive
black holes already identified.

\bigskip

\noindent
The LIGO discovery \cite{LIGO} of gravitational waves from black hole mergers
offers some support for our dark matter theory but it is premature to take this
support too seriously\cite{Riess}.

\bigskip

\section*{Discussion}

\bigskip

\noindent
In this article we have discussed three possibilities for
the solution of the dark matter problem although one of the three,
the first one we discussed, that involving the PIMBHs, is the most different. Unlike the others, 
it does not need to assume any new physics beyond the
standard model of particle theory. Instead, it relies on Einstein's
equations of general relativity, their black holes solutions and especially
the idea of entropy as developed in the nineteenth century.

\bigskip

\noindent
The other two possibilities are more similar to each other as
they both assume new physics beyond the standard model.
One (WIMP) assumes a supersymmetric extension of the electroweak
sector to ameliorate the scalar quadratic divergence problem
and the other (axion) assumes an extension of the
strong interaction QCD sector to solve the strong CP problem.
The WIMP and the axion particles are predicted in quite
different mass ranges and are the subjects therefore
of quite different ongoing experiments.
Certainly all such experiments are well worth pursuing.

\bigskip

\noindent
By assumption, the WIMP experiences weak interactions so it can
be searched for by direct detection of collisions with nuclei in the
laboratory, or by direct production in colliders like the LHC. Indirect
detection of WIMPs can be by searching for the products of WIMP
annihilation such as gamma rays and neutrinos in nearby galaxies
and clusters of galaxies. The WIMP is
expected in the mass range $10$ GeV to $1$TeV,

\bigskip

\noindent
The axion first appeared as a particle postulated to resolve the strong
CP problem of QCD. The original axion
was ruled out by experiment but was 
replaced in the theory
by a very-weakly-coupled very light ``invisible" axion. This was initially thought
to be undetectable until it was pointed out that one way to detect
such axions in the dark matter halo is by using 
cold resonant cavities with a magnetic field in which dark halo axions are converted into photons. 
The preferred mass for axions is in the range 1 ${\mu}$eV to 1 meV.

\bigskip

\noindent
The other solution for dark matter we discussed
is one which requires no new physics
but uses old physics dating from even before the birth of quantum field theory. 
In that theory, dark matter exists because 
Nature tried to maximise entropy in the early universe. This
was accomplished by producing Kerr black holes which concentrate
entropy many orders of magnitude more efficiently than anything else.
The masses of the most readily detectable primordial intermediate-mass black holes
(PIMBHs) are in the range from $25M_{\odot}$ to $2500M_{\odot}$.
It will be interesting to learn whether these PIMBHs show up
in the microlensing observations within the not-too-distant future.

\bigskip

\noindent
But there is a more fundamental, and we believe decisive, 
distinction between the three dark matter
solutions being discussed. It seems to us equally 
as important to understand
{\it why} dark matter exists, as it is to understand {\it what} it is.
Let us take as representative masses of the three candidates:
WIMP : 100 GeV ; Axion : 1 $\mu eV$ ; PIMBH : $100 M_{\odot}$. 
Given that the total mass of dark matter in the visible universe is
$\sim 10^{23} M_{\odot}$, the choice is between $10^{78}$ WIMPs,
$10^{95}$ axions or $10^{21}$ PIMBHs. For the first two cases 
there is no reason {\it why} dark matter exists.

\bigskip

\noindent 
For the case of PIMBHs, however, there is a clear reason
why so many were formed during the first three years after the Big Bang. That reason
is {\it entropy}. Entropy thus explains why about one quarter of the energy of the
universe is in the form of PIMBHs, a prediction which is soon readily testable.

\bigskip

\noindent
We have argued on the basis of Boltzmann's H theorem that entropy will increase
in an isolated out-of-equilibrium system as happens when the large numbers
of PBHs are formed in the first three years after the Big Bang. There is,
however, an apparent paradox because although the curvature of
spacetime, the gravitation, is undergoing strong fluctuations and
inhomogeneities, at the same time\cite{PHF999} the
electromagnetic photon-electron-proton plasma is in perfect thermal
equilibrium. One way out is simply to say that the gravitational and
electromagnetic sectors are decoupled, but this apparent paradox
merits further study,

\bigskip

\noindent
Nevertheless, we do know that black holes are by a very wide margin 
the best concentrators of entropy.
Therefore, the reason dark matter exists is as
foreseen over 150 years ago by the great physicist
Rudolf Clausius in a useful short statement of the second law 
of thermodynamics, {\it viz}.\\
the entropy of the universe tends to a maximum.

\bigskip

\noindent
\section*{Acknowledgements}

\noindent
We thank S. Altmann for useful discussions.
\end{document}